\documentclass[10pt, conference]{IEEEtran}

\ifCLASSOPTIONcompsoc

\usepackage[nocompress]{cite}
\else
 
\usepackage{cite}

\fi

\usepackage{amsthm}
\usepackage{graphicx}
\usepackage{float}
\usepackage{amssymb,amsmath}
\usepackage{subeqnarray}
\usepackage{cases}
\usepackage{epstopdf}
\usepackage{caption}
\usepackage{array}
\usepackage{algorithm}
\usepackage{algorithmic}
\usepackage{multirow}
\usepackage{xcolor}

\newtheorem{Definition}{Definition}
\newtheorem{Theorem}{Theorem}
\newtheorem{Lemma}{Lemma}
\usepackage{amsfonts,amssymb}
\usepackage[numbers]{natbib}
\usepackage{graphicx}
\begin{document}
	
\renewcommand{\figurename}{Fig.}

\title{Batch Auction Design For Cloud Container Services}

\author{\IEEEauthorblockN{Lin Ma}
	\IEEEauthorblockA{School of Computer\\
		Wuhan University\\
		linmawhu@gmail.com}
	\and
		\IEEEauthorblockN{Ruiting Zhou}
	\IEEEauthorblockA{Dept. of Computer Science\\
		The University of Calgary\\
		rzho@ucalgary.ca}
	\and
\IEEEauthorblockN{Zongpeng Li}
\IEEEauthorblockA{School of Computer\\
	Wuhan University\\
	zongpeng@whu.edu.cn}}
\maketitle
\begin{abstract}
Cloud containers represent a new, light-weight alternative to virtual machines in cloud computing. A user job may be described by a container graph that specifies the resource profile of each container and container dependence relations. This work is the first in the cloud computing literature that designs efficient market mechanisms for container based cloud jobs. Our design targets simultaneously incentive compatibility, computational efficiency, and economic efficiency. It further adapts the idea of batch online optimization into the paradigm of mechanism design, leveraging agile creation of cloud containers and exploiting delay tolerance of elastic cloud jobs. The new and classic techniques we employ include: (i) compact exponential optimization for expressing and handling non-traditional constraints that arise from container dependence and job deadlines; (ii) the primal-dual schema for designing efficient approximation algorithms for social welfare maximization; and (iii) posted price mechanisms for batch decision making and truthful payment design. Theoretical analysis and trace-driven empirical evaluation verify the efficacy of our container auction algorithms. 

\end{abstract}

\IEEEpeerreviewmaketitle

\section{Introduction}
Cloud computing offers cloud users with utility-like computing services on a pay-as-you-go fashion \cite{Qiu2012Cost}. Computing resources including CPU, RAM, disk storage and bandwidth can be leased in custom packages with minimal management overhead. %Cloud platforms on the internet today can be categorized into two groups. One is large-scale internet data centers (IDC), exemplified by Amazon EC2 \cite{a}, Microsoft Azure~\cite{azurecontainer} and Aliyun \cite{c}. These platforms organize a shared resource pool for serving cloud users. The other is co-location data centers \cite{d}, which gathers smaller clouds and provide unification management. 
Virtualization technologies help cloud providers pack cloud resources into a functional package for serving user jobs. Such packages used to be dominantly virtual machines (VMs), until the recent emergence of {\em cloud containers}, {\em e.g.}, Google Container Engine (largest Linux container) \cite{googlecontainer}, Amazon EC2 Container Service (ECS) \cite{amazonecs}, Aliyun Container Service \cite{Aliyuncontainer}, Azure Container Service \cite{azurecontainer}, and IBM Containers. Compared with general-purpose VMs, containers are more flexible and lightweight, enabling efficient and agile resource management. Applications are encapsulated inside the containers without running in a dedicated operating system \cite{f}. A representative cloud container is only megabytes in size and takes seconds to start \cite{f}, while launching a VM may take minutes. In the era of using VMs, VMs remain open throughout the life of the job. Because of the transient nature of a container, jobs could be seperated into several containers, and resource allocation is more convenient.

%and cloud providers may collect all the adaptive and flexible resources to adapt different jobs' requirement. 

%For example, as a vanguard in cloud services, Amazon is now one of the largest container service providers. In Amazon ECS, each cloud user submits a job definition including resource requirement, type of docker image, dependence graph of containers, and environment variables. ECS provisions the containers on a shared operating system, instead of running a virtual machine with a complete operating system. 

A complex cloud job in practice is often composed of {\em sub-tasks} \cite{Li2013Profit}. For example, a social game server \cite{e} typically consists of a front-end web server tier, a load balancing tier and a back-end data storage tier; a network security application may consist of an intrusion detection system (IDS), a firewall, and a load balancer. Different sub-tasks require different configurations of CPU, RAM, disk storage and bandwidth resources. Each sub-task can be served by a custom-made container following the resource profile defined by the cloud user \cite{He2012Elastic}. Some cloud containers are to be launched after others finish execution, following the input-output relation of their corresponding tasks. Such a dependence relation among containers is captured by a {\em container (dependence) graph}. For example, in Amazon ECS, a cloud user submits a job definition including resource requirements, type of docker image, a container graph, and environment variables. ECS then provisions the containers on a shared operating system, instead of running VMs with complete operating systems \cite{Tosatto2015Container}. 
 %\vspace*{-4mm}
%\begin{figure}[!htbp] 
 % \centering  
%  \includegraphics[width=80mm]{container.pdf}
%  \vspace*{-3mm}
%  \hspace*{-5mm}
%  \caption{Batch auction of cloud jobs running on containers.}  
%  \label{fig:1}  
%\end{figure} 

In the growing cloud marketplace ({\em e.g.}, Amazon EC2 and ECS), fixed pricing mechanisms \cite{Zhao2014Dynamic} and auctions complement each other. 
While the former is simple to implement, the latter can automatically discover the market price of cloud services, and allocate resources to cloud users who value them the most \cite{g}. 
A series of recent cloud auction mechanisms implicitly aim at non-elastic cloud jobs. These include both one-round cloud auctions \cite{g} and online cloud auctions \cite{h}, \cite{i}. In both cases, the provider processes each bid immediately and commits to an irrevocable decision. Furthermore, even in the online auctions, users' service time window is {\em predefined} by start and finish times in the bid \cite{h}, \cite{i}. %In order to meet the realistic situation, we should not only decide when to accept the bid but arrange reasonable schedule for each winner before its deadline. A simple illustration will help, one cloud user give a bid with cloud-container requirement, which will be executed in 2 hours, but the deadline prescribed in the bid is 36 hours. Why don't we wait for a period of time allow for there is plenty of time for us to make judgements, such as 20 minutes. Making a better judgement to maximize social efficiency according to the bids come between 20 minutes, and scheduling jobs within their tolerance time window. ({\bf rewrite: motivating batch cloud auctions})

A large fraction of cloud jobs are elastic in nature, as exemplified by big data analytics and Google crawling data processing. They require a certain computing job to be completed without demanding always-on computing service, and may tolerate a certain level of delay in bid acceptance and in job completion. For example, since Sanger {\em et al.} published the first complete genome sequence of an organism in 1977, DNA sequencing algorithms around the globe currently produce 15 billion gigabytes of data per annum, for cloud processing \cite{BiologyWatches}. A typical job of DNA testing takes $4$ hours to complete, while the user is happy to receive the final result anytime in a few days after job submission \cite{abc}. %An efficient cloud auction mechanism needs to judiciously exploit the tolerable time window from the user, through careful job scheduling. 

Given that bids from cloud users can tolerate a certain level of delay in bid admission, it is natural to revise the common practice of immediate irrevocable decision making in online cloud auctions. We can group bids from a common time window into a batch, and apply batch bid processing to make more informed decisions on all bids from the same batch simultaneously. Actually, if one considers only online optimization and not online auctions, then such batch processing has already been studied in operations research, such as online scheduling to minimize job completion time \cite{batchprocessing}, and scheduling batch and heterogeneous jobs with runtime elasticity in cloud computing platforms \cite{batchscheduling}.

We study efficient auctions for cloud container services, where a bid submitted by a cloud user specifies: (i) the container dependence graph of the job; (ii) the resource profile of each container; (iii) the deadline of the job; and (iv) the willingness to pay (bidding price). Cloud containers can be agilely created and dropped to handle dynamic sub-tasks in cloud jobs; it becomes practically feasible to suspend and resume a sub-task. As long as a container is scheduled to run for a sufficient number of time slots, its sub-task will finish.% Fig.~\ref{fig:1} illustrates the cloud job market based on auctions of container services. 

This work advances the state-of-the-art in the literature of cloud auctions along two directions. {\bf {\em First}}, while {\em batch algorithms} have been extensively studied in the field of online optimization, to the authors' knowledge, this work is the first that studies {\em batch auctions} in online auction design. {\bf{\em Second}}, this work is the first cloud auction mechanism designed for container services, with expressive bids based on container graphs. Our mechanism design simultaneously targets the following goals: (i) {\em truthfulness}, {\em i.e.}, bidding true valuation for executing its job on the cloud maximizes a user's utility, regardless of how other users bid; %Such a property is the holy grill in mechanism design, for it helps greatly simply the design and analysis of the resulting auction mechanism; 
(ii) {\em time efficiency}, we require that all components of the auction run in polynomial time, for practical implementation; (iii) {\em expressiveness}; the target auction permits a user to specify its job deadlines, desired cloud containers, and inter-container dependence relations; and (iv) {\em social welfare maximization}; {\em i.e.}, the overall `happiness' of the cloud-ecosystem is maximized. 

Corresponding to the above goals, our auction design leverages the following classic and new techniques in algorithm and mechanism design. 
For effectively expressing and handling user bids that admit deadline specification and container dependence graphs, we develop the technique of {\em compact exponential Integer Linear Programs (ILPs)}. We transform a natural formulation of the social welfare optimization ILP into a compact ILP with an exponential number of variables corresponding to valid container schedules. Although such a reformulation substantially inflates the ILP size, it lays the foundation for later efficient primal-dual approximation algorithm design, helping deal with non-conventional constraints that arise from container dependence and job deadlines, whose dual variables are hard to interpret and update directly. A combinatorial sub-routine later helps identify good container schedules efficiently without exhaustively enumerating them.
%This technique allows one to use both conventional and non-conventional linear constraints to express user requirements in the social welfare optimization problem, and subsequently combine techniques from linear optimization and combinatorial optimization in a unified algorithm framework. Here conventional constraints are ones that are well studied, whose dual variables are easy to understand and update in an iterative primal-dual optimization algorithm. Typical examples include packing and covering constraints \cite{primaldual}.
%(expressiveness - compact exponential)

Towards truthful batch auction design, we leverage the recent developments in {\em posted price auctions} \cite{Zhang2011Online}. At a high level, such an auction maintains an estimate of marginal resource prices for each resource type, based on expected supply-demand. Then upon decision making of each batch of bids, it chooses bids whose willingness to pay surpasses the estimated cost to serve them, based on resource demand of the container graph and projected marginal prices of resources. A winning user is charged with such estimated cost, which is independent from its bidding price. Truthfulness is hence guaranteed based on Myerson's celebrated characterization of truthful mechanisms \cite{myerson}.
%(online truthful auction - posted pricing)

The social welfare maximization problem in our container auction is NP-hard even in the offline setting, with all inputs given at once. A third key element of our cloud container auction is the classic primal-dual schema for designing efficient approximation algorithms, with rigorous guarantee on worst case performance. This is further integrated with the posted price framework, in that the marginal resource prices are associated with dual variables. 
The primal dual framework relies on a sub-routine that computes the optimal schedule of a given container graph, based on static resource prices (fixing dual variables, update primal solution). We apply dynamic programming \cite{Golin1998A} and graph traversal algorithms \cite{Maheshwar2012Bounding}, for designing the sub-routine for (i) service chain type jobs from network function virtualization, and (ii) general jobs with arbitrary topologies in their container graphs. We evaluate the effectiveness of our cloud container auction through rigorous theoretical analysis and trace-driven simulation studies. 

In the rest of the paper, we discuss related work in Sec.~\ref{related}, and introduce the auction model in Sec.~\ref{model}. The container auction is presented and analyzed in Sec.~\ref{topic} and Sec.~\ref{sec2} separately. Sec.~\ref{evaluation} presents simulation studies, and Sec.~\ref{sec:conclusion} concludes the paper.

\section{Related Work}\label{related}
%There exist a large body of studies in recent cloud computing literature on cloud auction design \cite{g,h,i,Zhou2016An}. Zaman {\em et al.} \cite{Combinatorial} study dynamic resource provisioning that takes into account user demands for VMs. Zhang {\em et al.} \cite{g} model the dynamic resource provisioning problem using an integer linear program, and design an approximation algorithm for social welfare maximization based on a primal-dual decomposition technique. Zhang {\em et al.} \cite{i} study online cloud auctions that model stochastic user arrivals in practice, and aim to optimize the packing of VMs with user-specified start and finish times. The above studies do not estimate resource or VM prices based on projected supply-demand in the near future. Furthermore, they all focus on auctions of VMs or bundles of unrelated VMs; in contrast, we target a newer, and mathematically richer, model of cloud jobs based on container graphs \cite{f}.

There exist a large body of studies in recent cloud computing literature on cloud auction design. Shi {\em et al.} \cite{rsmoa} studied online auctions where users bid for heterogeneous types of VMs and proposed RSMOA, an online cloud auction for dynamic resource provisioning. Zhang {\em et al.} \cite{onlineframework} propose COCA, a framework for truthfull online cloud auctions based on a monotonic payment rule and utility-maximizing allocation rule. These auction mechanisms are all confined to the solution space of immediately accepting or rejecting an arriving bid. To our knowledge, this work is the first that designs batch-type online auctions, both in the field of cloud computing and in the general literature of auction mechanism design. 
   
In terms of batch-type online algorithms, Deng {\em et al.} \cite{batchprocessing} study online scheduling in a batch processing system. Kumar {\em et al.} \cite{batchscheduling} design scheduling mechanisms for runtime elasticity of heterogeneous workloads. They propose Delayed-LOS and Hybrid-LOS, two algorithms that improve an existing dynamic programming based scheduler. These work possess a resemblance to ours in terms of postponing immediate response for more informed decision making, although they focus on algorithm design only and do not consider payments or incentive compatibility.
%Consequently we put forward a batch cloud auction using batch processing to complete a waiting-operation mechanism.  
%({\bf are there more studies on batch-type online optimization?})

Along the direction of posted price algorithms and mechanisms, Huang {\em et al.} \cite{Welfaremaximization} study online combinatorial auctions with production costs. They show that posted price mechanisms are incentive compatible and achieve optimal competitive ratios. Etzion {\em et al.} \cite{postprice} present a simulation model to extend previous analytical  framework, focusing on a firm selling consumer goods online using posted price and auction at the same time. This work was inspired in part by this line of recent developments on using posted prices to achieve effective resource allocation and bid-independent charges.

%The primal dual framework has been recently applied in the design of a series of online auctions in power demand response \cite{smartgrids} and in cloud computing \cite{Welfaremaximization}. All these studies design traditional online auctions with immediate irrevocable decision making upon the arrival of each bid. To our knowledge, this work is the first that designs batch-type online auctions, both in the field of cloud computing and in the general literature of auction mechanism design. 
%In this work, we make decisions after a short time slots, it helps us make more accurate judgement for cloud users and better forecast for the future. This work is first to design a container based batch auction to handle. 

\section{The Cloud Container Auction Model}\label{model}
%In a public cloud, the cloud provider is responsible to manage available resources in its data center. The provider accepts requests in the form of {\em job bids} from cloud users. If a bid is accepted, the provider encapsulates resources into tailored {\em containers} to serve that job. %In designing the container-based auction for the cloud market, our goal is to maximize resource utilization for maximizing the social welfare of the cloud ecosystem, which further requires judicious scheduling the tasks embedded in each accepted job. 

We consider a public cloud in which the cloud provider (auctioneer) manages a pool of {\em R} types of resources, as exemplified by CPU, RAM, disk storage and bandwidth,  %that can be dynamically packed into different cloud containers. 
and the capacity of resource-{\em r} is {\em $\mathbb{C}_{r}$}. Integer set \{1, 2,..., X\} is denoted by [{\em X}]. %Thanks to the lightweight property of containers, scheduling jobs in non-consecutive time slots (suspending and then resuming a job) becomes practically viable, as compared to full-fledged virtual machines. %as long as the constraint of resource-{\em r} is satisfied. Then we have this basic model.
There are {\em I} cloud users arriving in a large time span \{1, 2, ..., T\}, acting as bidders in the auction. %While the users arrive in a stochastic process, our auction design does not depend on a specific assumption on the process ({\em e.g.}, Poisson arrival). 
Each user $i$ submits a job bid that is $4$-tuple:

\vspace*{-3mm}
{\small
\begin{equation}
\Pi_{i} = \{\mathcal{W}_{i}, t_{i}, d_{i}, B_{i}\}.
\end{equation}   
}
Here $\mathcal{W}_{i}$ is the workload of user $i$, $t_{i}$ is arrival time of user {\em i}, and its required deadline for job completion is $d_{i}$. $B_{i}$ is user $i$'s overall willingness-to-pay for finishing its job by $d_i$. 

According to users workload, the detailed information will be obtained by cloud platform. Such as the number of sub-tasks of the job {\em M}, and each sub-task requires a container to process, thus {\em m} is also the number of containers. The container graph $G_{i}$ that describes the dependence among sub-tasks. The number of requested time slots for each sub-task $N_{im}$. Each sub-task can be suspended and resumed, as long as the total execution time accumulates to $N_{im}$. $h^{r}_{im}$ is the resource configuration of container{\em m} of user {\em i}. 

%\begin{figure}[!htbp]
%	\begin {center}
%	\includegraphics[width=0.50\textwidth]{JobContainer.pdf}	
%	\vspace{-1mm}
%	\caption{Container graphs for cloud jobs from \cite{containermodel}.}
%	\label{fig:Model}
%	\end {center}
%	\vspace*{-4mm}
%\end{figure}

A {\em (container) schedule} is a mapping from resources and time slots to cloud containers, serving accepted cloud jobs to meet their deadlines. 
%In existing literature of online cloud auction design, a decision to accept/reject a bid is made immediately upon its submission, despite the fact that a cloud job can often tolerate a certain threshold of delay before receiving the decision. 
We postpone immediate decision making on the bids, to judiciously exploit cloud jobs' tolerable delays in bid admission. %We grouping incoming cloud bids into batches, making decisions on each batch simultaneously, targeting more informed decision making for higher social welfare. More specifically, w
We group bids from every $\theta$ time slots into a batch, resulting in {\em Q} batches within the large time span {\em T}. Let $\rho_{q}$ be the number of users arriving within batch $q \in Q$. A binary variable $x_{i}$ indicates whether user {\em i}'s bid is accepted (1) or not (0). Another binary variable $z_{im}(t)$ indicates whether to execute user $i$'s sub-task $m$ at time slot {\em t} (1) or not (0); it encodes a schedule of user $i$'s job. The cloud provider further computes a payment $P_{i}$ to charge for a winning cloud user $i$. The holy grail of auction mechanism design is {\em truthfulness}, the property that greatly simplifies bidder strategy space and analysis of the auction mechanism.  

\begin{table}[!htbp] 
\begin{center}
\caption{ List of Notations }
\begin{tabular}{|c|l|l|l|}
\hline
I&\multicolumn{3}{|l|}{$\#$ of users}\\
\hline
T&\multicolumn{3}{|l|}{$\#$ of time slots}\\
\hline
$\mathbb{C}_{r}$&\multicolumn{3}{|l|}{capacity of type-$r$ resource}\\
\hline
M&\multicolumn{3}{|l|}{$\#$ of sub-tasks/containers of one job}\\
\hline
$\mathcal{W}_{i}$&\multicolumn{3}{|l|}{workload of user $i$}\\
\hline
$G_{i}$&\multicolumn{3}{|l|}{dependence graph of user {\em i}'s sub-tasks}\\
\hline
$N_{im}$&\multicolumn{3}{|l|}{$\#$ of time slots requested by user {\em i}'s container {\em m}}\\
\hline
$h_{im}^{r}$&\multicolumn{3}{|l|}{demand of type-{\em r} resource by user {\em i}'s  container {\em m}}\\
\hline
$t_{i}$&\multicolumn{3}{|l|}{user {\em i}'s arrival time}\\
\hline
$d_{i}$&\multicolumn{3}{|l|}{deadline of user {\em i}'s bid}\\
\hline
$B_{i}$&\multicolumn{3}{|l|}{bidding price of user {\em i}'s bid}\\
\hline
$x_{i}$&\multicolumn{3}{|l|}{accept the user {\em i}'s bid(1) or not(0)}\\
\hline
$\rho_{q}$&\multicolumn{3}{|l|}{$\#$ of users arriving within  batch $q$}\\
\hline
$f_{ir}^{S}(t)$&\multicolumn{3}{|l|}{total type-$r$ resource occupation of schedule in $\Gamma_{i}$ for slot {\em t}}\\
\hline
$\theta$&\multicolumn{3}{|l|}{$\#$ of time slots within one batch interval }\\
\hline
$z_{im}(t)$&\multicolumn{3}{|l|}{allocated user {\em i}'s container {\em m} at time slot {\em t}(1) or not(0)}\\
\hline
$w_{r}(t)$&\multicolumn{3}{|l|}{amount of allocated type-$r$ resource at time {\em t}}\\
\hline
$y_{r}(t)$&\multicolumn{3}{|l|}{availablity of type-{\em r} resource at time slot {\em t}}\\
\hline
$\kappa_{r}(t)$&\multicolumn{3}{|l|}{marginal price of type-{\em r} resource at time slot {\em t}}\\
\hline
$F_{r}$&\multicolumn{3}{|l|}{ minimum value of user's valuation per unit of type-r resource}\\
\hline
$D_{r}$&\multicolumn{3}{|l|}{ maximum value of user's valuation per unit of type-r resource}\\
\hline
$\Gamma_{i}$&\multicolumn{3}{|l|}{ the set of valid schedules for each user}\\
\hline
$u_{i}$&\multicolumn{3}{|l|}{ user {\em i}'s utility}\\
\hline

\end{tabular}
\end{center}
\end{table}

%
%\begin{Definition}(Truthfulness in bidding price)
% An auction mechanism is {\em truthful} if a user $i$ always maximizes its utility by submitting true valuation $\upsilon_{i}$ for its job, regardless of other users' bids.% every user's valuation is never under the influence of other user's bidding price.
%\end{Definition}

\begin{Lemma}\label{myerson}
	Let $Pr(B_{i})$ denote the probability of bidder $i$ winning an auction and $B_{-i}$ be the bidding price except $i$. A mechanism is truthful if and only if the following hold for a fixed $B_{-i}$ {\em \cite{Gopinathan2011Strategyproof}}:
	
	1) $Pr(B_{i})$ is monotonically non-decreasing in $B_{i}$;
	
	2) bidder $i$ is charged by $B_{i}Pr(B_{i}) - \int_{0}^{B_{i}} Pr(B_{i})dB$.
\end{Lemma}

Lemma \ref{myerson} can be explained in this orientation: the payment charged to bidder {\em i} for a fixed $B_{i}$ is independent of $B_{i}$. We will use this mode to design a posted price function in Sec.~\ref{topic}. Since we meet the challenge that when we consider that online batch auction decisions are to be made based on hitherto information only.
%Let the {\em true valuation} of user $i$'s job be $\upsilon_{i}$. 
If user $i$'s job is accepted, its utility is $u_{i} = \upsilon_{i}-P_{i}$, which equals $u_{i} = B_{i}-P_{i}$ under truthful bidding. The cloud provider's utility is $\sum_{i\in[I]}P_{i}$. The {\em social welfare} that captures the overall utility of both the provider and the users is ($\sum_{i\in[I]}B_{i}x_{i}-\sum_{i\in[I]}P_{i}$) + ($\sum_{i\in[I]}P_{i}$). With payments cancelling themselves, the social welfare is simplified to $\sum_{i\in[I]}B_{i}x_{i}$.

Under the assumption of truthful bidding, the Social Welfare Maximization problem in our cloud container auction can be formulated into the following Integer Linear Program (ILP):

\vspace*{-2mm}
{\small
\begin{align*}
\mbox{maximize}\quad \sum_{i\in[I]} B_{i}x_{i}  \tag{2} \label{max1}
\end{align*} 
subject to:

\vspace*{-3mm}
\begin{subequations}
\begin{equation}
\begin{aligned}
\theta \lceil \frac{t_{i}}{\theta}\rceil x_{i}  &\leq tz_{im}(t),\forall t, \forall m, \forall i:t_{i}\leq t,\label{m1a}
\end{aligned}
\end{equation}
\begin{equation}
\begin{aligned}
tz_{im}(t)&\leq  d_{i}x_{i} ,\forall t, \forall m, \forall i:t_{i}\leq t, \label{m1b} 
\end{aligned}   
\end{equation}
\begin{equation}
\begin{aligned}
tz_{im}(t)&\leq t'z_{im'}(t'), 
\label{m1f}
\end{aligned}
\end{equation}
\begin{equation*}
\forall t, t', 
\forall i: task\ m'\ \textrm{arrives later}\ than\ task\ m,
\end{equation*}
\begin{equation}
\begin{aligned}
N_{im}x_{i}&\leq \sum_{t\in [T]}z_{im}(t), \forall m, \forall i,\label{m1c}
\end{aligned}
\end{equation}
\begin{equation}
\begin{aligned}
\sum_{i\in[I]} \sum_{m\in[M]} h_{im}^{r} z_{im}(t)&\leq  \mathbb{C}_{r}, \forall r, \forall t, \label{m1d}
\end{aligned}
\end{equation}
\begin{equation}
 x_{i},z_{im}(t) \in {\{0,1\}}, \forall i,\forall t,\forall m.
 \label{m1e}
 \end{equation}
\end{subequations}}
Constraints (\ref{m1a}) and (\ref{m1b}) ensure that user $i$'s job is scheduled to execute only between its start time and deadline. (\ref{m1f}) enforces inter-task dependence of user {\em i}'s sub-tasks, and (\ref{m1c}) makes sure that the total number of allocated time slots for each container is sufficient to finish the corresponding sub-task. Constraint (\ref{m1d}) states that the total amount of type-$r$ resource utilized at time slot $t$ is capped by system capacity. %Finally, each bidder can win at most 1 bid at one time, it is illustrated with (2e).\\

Even in the offline setting with all inputs given, ILP (\ref{max1}) is still NP-hard. This can be verified by observing that with constraints (2e) and 2(f) alone, and ILP (\ref{max1}) degrades into the classic knapsack problem known to be NP-hard. We resort to the classic primal-dual schema \cite{primaldual} for efficient algorithm design. We first reformulate ILP (\ref{max1}) into an equivalent {\em compact exponential} version, to hide the non-conventional constraints that arise from container dependence and job deadlines, whose dual variables would be hard to interpret and to update:
%\vspace*{-3mm}
{\small 
\begin{align*}
\mbox{maximize}\quad \sum_{i\in[I]} \sum_{S\in\Gamma_{i}}B_{i}x_{iS}   \tag{3}\label{max2}
\end{align*}
subject to:

\vspace*{-3mm}
\begin{subequations}
\begin{equation}
\sum_{i\in[I]}  \sum_{S:t\in S} f_{ir}^{S}(t) x_{iS}\leq  \mathbb{C}_{r}, \forall r\in[R], \forall t\in[T], \label{m2a}
\end{equation}
\begin{equation}
\sum_{S\in \Gamma_{i}}x_{iS} \leq 1, \forall i\in[I],\label{m2b}
\end{equation}
\begin{equation}
x_{iS} \in {\{0,1\}} ,\forall i\in[I],\forall S\in \Gamma_{i}.\label{m2c}
\end{equation}
\end{subequations}}
In the compact exponential ILP above, $\Gamma_{i}$ represents a set of valid schedules for sub-tasks that meet constraints (\ref{m1a}), (\ref{m1b}), (\ref{m1f}) and (\ref{m1c}). $B_{iS}$ represents the bidding price of user $i$ in schedule $S \in\Gamma_{i}$. Since a time slot can serve two or more containers, we let $f_{ir}^{S}(t)$ represent the total type-$r$ resource occupation of user {\em i}'s schedule {\em S} in {\em t}. Constraints (\ref{m2a}) and (\ref{m2b}) correspond to (\ref{m1d}) and (\ref{m1e}) in ILP (\ref{max1}). We relax the integer constraints $x_{i} \in {\{0,1\}}$ to $x_{i} \geq 0$, and introduce dual variable vectors $u_{i} $ and $ \kappa_{r}(t)$ to constraints (\ref{m2a}) and (\ref{m2b}) respectively, to formulate the dual of the LP relaxation of ILP (\ref{max2}). %This pair is to be used in the primal-dual algorithm design in next section:

\vspace*{-3mm}
{\small
\begin{align*}
 \mbox{minimize} \sum_{i \in [I]} u_{i} + \sum_{t\in [T]} \sum_{r\in [R]}\mathbb{C}_{r}\kappa_{r}(t) \tag{4}\label{min}
\end{align*}
subject to:

\vspace*{-3.7mm}
\begin{subequations}
\begin{equation}
u_{i} \geq  B_{i}-\sum_{r \in[R]} \sum_{ t \in S} f^{S}_{ir}(t) \kappa_{r}(t), \forall i \in [I], \forall S\in \Gamma_{i} ,\label{m3a}
\end{equation}
\begin{equation}
\kappa_{r}(t),u_{i} \geq 0,\forall i \in [I], \forall r \in [R],\forall t \in [T].\label{m3b}
\end{equation}
\end{subequations}}
While the reformulated ILP (\ref{max2}) is compact in its form, it has an exponential number of variables that arise from the exponential number of feasible job schedules. Correspondingly, the dual problem (\ref{min}) has an exponential number of constraints. 
Even there are exponential number of schedule options are available, we only select polynomial number of them to compute the approximately optimal objective through a sub-algorithm (sec \ref{subalg}).
We next design an efficient auction algorithm that efficiently solves the primal and dual compact exponential ILPs simultaneously, pursuing social welfare maximization (in the primal solution) while computing payments (in the dual solution).

\section{BATCH AUCTION ALGORITHM FOR SOCIAL WELFARE MAXIMIZATION}\label{topic}

%({\bf MAIN PROBLEM: structure, logic flow, and content are UNCLEAR})

\subsection{The Batch Algorithm }
Departing from traditional online auctions that make immediate and irrevocable decisions, our auction mechanism takes a batch processing approach to handle user bids. In each batch, we aim to choose a subset of bids to accept, and to dynamically provision containers, through choosing a feasible assignment of the primal variable $x_{iS}$. We let $x_{iS} = 1$, if user {\em i}'s bid with schedule {\em S} is accepted, then allocate time slots according to the schedule, and update the amount of  resources occupied. % $w_{r}(t)$ by $f_{ir}^{S}(t)$ in {\em t}. 

We now focus on batch bid processing and container provisioning for social welfare maximization. A set of dual constraints exists for each primal variable $x_{iS}$. We minimize the increase of the dual objective and maintain dual feasibility (\ref{m3a}) by leveraging {\em complementary slackness}. Once the dual constraint (\ref{m3a}) is tight with user $i$'s schedule $S$ (KKT conditions \cite{i}), the primal variable $x_{iS}$ is updated to $1$. According to constraint (\ref{m3b}), the dual variable $u_{i} \geq 0$. Therefore, we let $u_{i}$ be the maximum of $0$ and the RHS of (\ref{m3a}). If $u_{i} = 0$, the bid is rejected.
{\small
\begin{equation}
u_{i} = \max\{0,\max_{s\in\Gamma_{i}}( B_{i}-\sum_{r \in[R]} \sum_{ t \in S} f^{S}_{ir}(t) \kappa_{r}(t))\} \label{equ:u},  \forall i \in \rho_{q}
\end{equation}}
$\kappa_{r}(t)$ can be viewed as the marginal price per unit of type-{\em r} resource at {\em t}. Consequently, {\small $\sum_{r \in[R]} \sum_{ t \in S} f^{S}_{ir}(t) \kappa_{r}(t))$} represents the cost of serving user {\em i} by schedule {\em S}, and $\{ B_{i}-\sum_{r \in[R]} \sum_{ t \in S} f^{S}_{ir}(t) \kappa_{r}(t)\}$ is the utility of user {\em i}'s bid. The above assignment (\ref{equ:u}) chooses the schedule which can maximize the job's utility..

Our auction strives to reserve a certain amount of resource for potential high-value bids in the future. Careful implementation of such an intuition through dual price design is crucial in guaranteeing a good competitive ratio of the auction. %Due to the dual price design we can schedule user's bid to meet his maximization utility. Below we present the marginal price design and details of batch processing.  

Let $D_{r}$ and $F_{r}$ represent the maximum and minimum user valuation per unit of type-{\em r} resource respectively. %The value of type-$r$ resource is bounded by users' $D_{r}$ and  $F_{r}$.  
$w_{r}(t)$ denotes the amount of allocated type-{\em r} resource at {\em t}. We define the marginal price $\kappa_{r}(t)$ to be an increasing function of $w_{r}(t)$: 
  {\small   
	\begin{equation}
	\kappa_{r}(w_{r}(t)) = \frac{\sigma F_{r}}{ k}(\frac{kD_{r}}{\sigma F_{r}})^\frac{w_{r}(t)}{\mathbb{C}_{r}} \label{equ:k}
	\end{equation}
	\begin{center}
		where $ D_{r}$ =  $ \max\limits_{i\in [I]}\frac{B_{i}}{\sum\limits_{m\in [M]}N_{im}h_{im}^{r}} $;	
		$  F_{r}$ = $\min\limits_{i\in [I]}\frac{B_{i}}{\sum\limits_{m\in [M]}N_{im}h_{im}^{r}} $.
	\end{center}}
The initial price of each type-{\em r} resource should be low enough such that any user's bid can be accepted; otherwise there might be a large amount of idle resource. Thus we decrease the starting price by a coefficient {\em k}, satisfying: $k-1 = \max_{r\in[R]}ln(\frac{k D_{r}}{\sigma F_{r}})$ and $ k>1 $. The detailed explanation of {\em k} is given in Theorem \ref{lem4}. For all $w_{r}(t) < \mathbb{C}_{r}$ , $\kappa_{r}(t) < D_{r}$, and it will reach $D_{r}$ when $w_{r}(t)$ = $\mathbb{C}_{r}$. In that case, the cloud provider will not further allocate any type-$r$ resource. %In summary, $\kappa_{r}(t)$ is defined as a function on $w_{r}(t)$ as follows: 
The parameter is defined as the minimum occupation rate of all kinds of resources within slots {\em T}, {\em i.e.},
{\small
\begin{center}
	$\sigma = \min_{r\in R}\frac{\sum_{i\in[I]}\sum_{m\in [M]}h_{im}^{r}N_{im}x_{i}}{\mathbb{C}_{r}T}$ 
\end{center}}
We assume that there are enough cloud users to potentially exhaust resources within each slot. Thus the resource occupation rate $\sigma$ is close to 1. %For a good competitive ratio, we decrease the starting price by a coefficient of {\em k}, which is satisfies the equation: $k-1 = ln(\frac{k D_{r}}{\sigma F_{r}}), k>1 $, the detailed explanation of {\em k} is given in Lemma \ref{lem:1}.  %(Due to the truthfulness of our auction mechanism, $\upsilon_{i}$ equals to $B_{i}$) ({\bf revise}):

%When each user's schedule is determined, let's consider the value of a unit resource to each user. $\sum_{r\in[R]}\sum_{t\in [s_{i}]}f_{ir}^{s}(t)\kappa_{r}(t)$ is viewed as the weighted total resource demand by user {\em i}'s job, thus 
%$\frac{B_{i}}{\sum_{r\in[R]}\sum_{t\in[s_{i}]}f_{ir}^{s}(t)\kappa_{r}(t)}$ can be interpreted as the value for a unit-weight resource for user {\em i}. In a round of batch processing, we first select the best user {\em i}'s job with the maximum unit resource value. By doing so, we revised the FCFS mechanism in an online auction mechanism into one that achieves a higher social welfare for the cloud system, through more informed and more intelligent bid selection.

We design a batch auction algorithm $A_{batch}$ in Alg.~\ref{alg:1} with container scheduling algorithm $A_{sub}$ in Alg.~\ref{alg:4} or Alg.~\ref{alg:2},which can select optimal container scheduling under different circumstances. 
%We design a batch auction algorithm $A_{batch}$ in Alg .~\ref{alg:1} with container scheduling algorithm $A_{sub}$ in Alg.~\ref{alg:4} and Alg.~\ref{alg:2} respectively.
$A_{batch}$ defines the posted price function and initializes the primal and dual variables in line 1. Upon the arrival of $\rho_{q}$ users within batch $q$, we first select the schedule that maximize users' utility through the dual oracle(lines 4-6).  $\sum_{r\in[R]}\sum_{t\in [s_{i}]}f_{ir}^{s}(t)\kappa_{r}(t)$ in line 7 is viewed as the weighted total resource demand by user {\em i}, thus $\frac{B_{i}}{\sum_{r\in[R]}\sum_{t\in[s_{i}]}f_{ir}^{s}(t)\kappa_{r}(t)}$ can be interpreted as the value for a unit resource of user {\em i}, and we select the bid $\mu$ with the maximum unit resource value. If user $\mu$ obtains positive utility, we update the primal variable $x_{\mu}$ and dual variable $\kappa_{r}(t)$ according to $\mu$'s schedule $s_{\mu}$ (lines 9-16).

\begin{algorithm}[htp]         
	\caption{ A Primal-dual Posted Price Auction $A_{batch}$ }             
	\label{alg:1}                

{\small
	\begin{algorithmic}[1]  
		          
		%	\REQUIRE bidding language {\bf $\{\Pi_{i}\}$}, \{$\mathbb{C}_{r}$\}
			%\STATE {\bf Initialization}
			
			\STATE   Initialize  $x_{i}=0$, $z_{im}(t)=0$, $w_{r}(t)=0$, $u_{i}=0$, $\kappa_{r}(t)=\frac{\sigma F_{r}}{k}$, $ \forall i \in [I], r\in [R], t \in [T], S\in \Gamma_{i}, \psi = \varnothing$; 
			%\STATE   Define $\kappa_{r}(t)$ according to (6);\\
			\STATE {\bf Group a set of $\rho_{q}$ users within $\theta$ time slots};
			\WHILE    {$\psi \neq \rho_{q}$}
			\FORALL {$i \in \rho_{q}\setminus\psi$}
			\STATE ($u_{i}$,$S_{i}$,$cost_{i}$,$\{f_{ir}^{S}(t)\}$)=$A_{sub}$($\{\Pi_{i}\}$,\{$\mathbb{C}_{r}$\},\{$w_{r}(t)$\}, \{$\kappa_{r}(t)$\});
			\ENDFOR
			\STATE $\mu = argmax_{i \in \rho_{q}\setminus\psi }$\{$\dfrac{B_{i}}{\sum_{r\in[R]}\sum_{t\in [s_{i}]}f_{ir}^{s}(t)\kappa_{r}(t)} $\};
			%\STATE \textbf{then} $x_{\mu}=1$;
			\IF  {$u_{\mu}>0$}
			\STATE  $x_{\mu}=1$;
			\STATE Accept user $\mu$'s bid, allocate resources according to $S_{i}$, and charge $cost_{i}$ for user {\em i};
			\STATE \textbf{update:} $ \psi = \psi \bigcup \{\mu \}$;
			\FORALL {$t \in S_{\mu}$}
			% \STATE $ w_{r}(t)=w_{r}(t)+f_{\mu r}^{s}(t), \forall r\in [R]$;
			\STATE $w_{r}(t) = w_{r}(t) + f^{S}_{\mu r}(t)$;
			\STATE $\kappa_{r}(t) = \frac{\sigma F_{r}}{ k}(\frac{ kD_{r}}{\sigma F_{r}})^\frac{w_{r}(t)}{\mathbb{C}_{r}},  \forall r\in [R]$;
		
			\ENDFOR
			\ELSE
			\STATE Reject user $\mu$'s bid, and delete user $\mu$ from the set $\rho_{q}$.
			\ENDIF
			\ENDWHILE     
	\end{algorithmic}}
\end{algorithm}

\subsection{Sub-algorithm of Auction Mechanism}\label{subalg}

Our container scheduling algorithms $A_{sub}$ only selects utility-maximizing schedules for each job, rather than an exponential number of schedules. Therefore, we compute a schedule that minimizes the cost of serving the job.

In our auction mechanism, dependence graph of user tasks is complicated to handle. We first focus on a relatively small, yet representative case of jobs from Network Function Virtualization \cite{nfv}, where each container graph is a service chain. We exploit the sequential chain structure to design $A_{sub1}$ Algorithm \ref{alg:4} with polynomial time complexity, based on dynamic programming. By choosing time slots that can ensure right operating sequence and minimum payment for each sub-task, the first two nested {\tt for} loops select minimum-cost schedule for containers (lines 3-10). Then the second {\tt for} loop updates the cost and schedule for each container {\em m} (lines 11-15); line $17$ updates the cost and utility of user {\em i}'s schedule $S_{i}$ at the end.

{\small
\begin{algorithm}[htb]        
	\caption{$A_{sub1}$: Container Graph Scheduling - Service Chains}             
	\label{alg:4}                  
{\small	\begin{algorithmic}[1]   
		
		\REQUIRE                           
		bidding language  $\{\Pi_{i}\}$, \{$\mathbb{C}_{r}$\},\{$\kappa_{r}(t)$\}, \{$w_{r}(t)$\};
		\ENSURE
		$ u_{i}$; $S_{i}$, $cost_{i} $,   $\{f^{S}_{ir}(t)\}$;
		\STATE Initialize $S_{i} = \varnothing$; $f^{S}_{ir}(t)=0,\forall t\in[T]$; 
		\FORALL  {$m\in[M]$}
		\FORALL {$t_{s} \in [\theta \lceil \frac{t_{i}}{\theta}\rceil+\sum_{1}^{m-1}N_{im},d_{i}-\sum_{m}^{M}N_{im}]$}
		\FORALL {$t_{e} \in [t_{s}+N_{im},d_{i}-\sum_{m+1}^{M}N_{im}]$}
		
		\STATE $c_{m}(t)=\sum_{r\in[R]}h_{im}^{r}\kappa_{r}(t),  \forall t\in [t_{s},t_{e}]$;
		\STATE Select $N_{im}$ slots with minimum $c_{m}(t)$ and $w_{r}(t)+h^{r}_{im} \leq \mathbb{C}_{r}, \forall r \in [R]$ to $ \tau_{m} $; 
		\STATE $\Delta_{m}=[\Delta_{m} \ \tau_{m}]$;
		\STATE  $p_{m}(t_{s},t_{e})$=$\sum_{t\in \tau_{m}} c_{m}(t)$;
		\ENDFOR
		\ENDFOR
		%\IF  {$m > 1$}
		\FORALL  {$t_{s} \in [\theta \lceil \frac{t_{i}}{\theta}\rceil+\sum_{1}^{m-1}N_{im},d_{i}-\sum_{m}^{M}N_{im}]$}
		\STATE $pay = min_{t_{e}<t_{s}}\{p_{m-1}(:,t_{e})\}, \tau_{m}\in [\Delta_{m}]$;
		\STATE  $p_{m}(t_{s},t_{e}) = p_{m}(t_{s},t_{e}) + pay$;
		\STATE  $S_{i}= [S_{i}\ \tau_{m}],\ f_{ir}^{S}(t) = f_{ir}^{S}(t)+h_{im}^{r}$;
		\ENDFOR
		%\ENDIF
		\ENDFOR
		\STATE {\bf Update:} $cost_{i} = min_{t_{s},t_{e}}(p_{m}(t_{s},t_{e}))$; $ u_{i}=B_{i}-cost_{i}$;

	\end{algorithmic}}
\end{algorithm}
}

Container graphs in practice can be more complex than a chain structure. For general jobs with arbitrary container graph topology, the container scheduling problem is NP-hard, as proven in Theorem.~\ref{themnp}; we design $A_{sub2}$ in Algorithm \ref{alg:2} to solve the optimization. Lines 2-8 in Algorithm \ref{alg:2} sort available time slots by $c_{m}(t)$. Then $A_{sub2}$ employs Depth-First Search (DFS) (line 9). We adapt the DFS procedure with improvements to select available time slots with minimum cost in a recursive process that decides a container schedule. Truthfulness requires solving the problem exactly, and our algorithm runs in exponential time to the number of sub-tasks in a job, which is mostly small and can be viewed as a constant in practice. 

\begin{Theorem}\label{themnp}
	In each batch of container based auction, given fixed resource prices, choosing the schedule of sub-tasks with minimum cost with a general container graph is NP-hard. 
	
\end{Theorem}
\noindent{\em Proof}:
We construct a polynomial-time reduction to sub-task scheduling from the classic NP-hard problem {\em subset sum}:
	$\max_{x_{i}}\sum_{i=1}^{n} c_{i}x_{i},
	\mbox{subject to} \sum_{i=1}^{n} c_{i}x_{i}\leq V, x_{i}\in \{0,1\}.$
	
Given a set $\{c_{1},c_{2},...,c_{n}\}$ and a objective {\em V}, our problem reduces to an instance of $K$ = ($|M| = n,h_{im}^{r}= c_{i}, \mathbb{C}_{r}=V$), in which each user's job has {\em M} types of containers with $1$ slot requirement, and the resource pool contains one type of resource. We should put as many containers in one slot with lowest price as possible. If a polynomial-time algorithm solves the capacitated container scheduling problem $K$, it will solve the corresponding subset sum problem as well, and vice versa. Consequently, the {\em subset sum} problem can be viewed as a special case of the sub-task scheduling problem, which must be NP-hard as well.
\qed

{\small
\begin{algorithm}[htb]        
	\caption{ $A_{sub2}$: Container Graph Scheduling - General Topology  }             
	\label{alg:2}                  
{\small	\begin{algorithmic}[1]   
		
		\REQUIRE                           
		bidding language $\{\Pi_{i}\}$, \{$\mathbb{C}_{r}$\},\{$\kappa_{r}(t)$\}, \{$w_{r}(t)$\};
		\ENSURE
		$ u_{i}$; $S_{i}$, $\{f^{S}_{ir}(t)\}$, $cost_{i}$;
		\STATE Initialize $S_{i} = \varnothing$; $f^{S}_{ir}(t)=0,\forall t\in[T]$; $c_{min}$=INF;
		\FORALL  {$m\in[M]$}
		\FORALL {$t\in [\theta \lceil \frac{t_{i}}{\theta}\rceil,d_{i}]$}
		\STATE $c_{m}(t)=\sum_{r\in[R]}h_{im}^{r}\kappa_{r}(t),  \forall t\in [\theta \lceil \frac{t_{i}}{\theta}\rceil,d_{i}]$;
		\STATE Sort slots with $w_{r}(t)+h^{r}_{im} \leq \mathbb{C}_{r}, \forall r \in [R]$ according to $c_{m}(t)$ to $ \tau_{m} $;
		\STATE  $p_{m}(t_{s},t_{e})$=$\sum_{t\in \tau_{m}} c_{m}(t)$; 
		\ENDFOR
		\ENDFOR
		
		\STATE Calling {\bf Depth-First Search(m)} to find the container schedule $S_{i}$ and resource allocaton $\{f^{S}_{ir}(t)\}$ with minimum cost $c_{min}$;
		\STATE \textbf{Update:} $cost_{i} = c_{min}$;$ u_{i}=B_{i}-cost_{i}$;

	\end{algorithmic} }
\end{algorithm}}

\iffalse 
\begin{algorithm}[htb]        
	\caption{ Sub-tasks Scheduling }             
	\label{alg:3}                  
	\begin{algorithmic}[1]  
		{\small
			\STATE  CSD(m)\{
			\IF  {m == M}
			
			\IF  {$c < c_{min}$ and $u_{i}=B_{i}-c>0$}
			\STATE $c_{min} = c$; $s_{i}(m,N_{im})$=$L(m,N_{im})$;
			\ENDIF
			\STATE  Return;
			\ELSE
			\STATE  Delete slot {\em t} from $\tau_{m}$ if $w_{r}(t)+h^{r}_{im} < \mathbb{C}_{r}, \forall r \in [R]$; 
			\IF {$length(\tau_{m}) < N_{im}$}
			\STATE  Return; 
			\ELSE 
			\STATE Select $N_{im}$ feasible slots from $\tau_{m}$ to $S_{mJ}$;
			\FORALL {$S_{mj}, \forall j\in[J]$}
			\STATE $f^{S}_{ir}(t) = f^{S}_{ir}(t) + h_{im}^{r}, \forall t\in S_{mj}$;
			\STATE Put all the time slots from $S_{mj}$ in $L(m,N_{im})$;
			
			\STATE $c = c + \sum_{t\in S_{mj}}\sum_{r}\kappa_{r}(t)h_{im}^{r}$;
			\IF {$c>c_{min}$}
			\STATE Return;
			\ENDIF
			\STATE CSD(m+1);
			\STATE $c = c - \sum_{t\in S_{mj}}\sum_{r}\kappa_{r}(t)h_{im}^{r}$;
			\STATE $f^{S}_{ir}(t) = f^{S}_{ir}(t) - h_{im}^{r}, \forall t\in S_{mj}$;
			\STATE Delete all the time slots from $S_{mj}$ in $L(m,N_{im})$;
			\ENDFOR
			\ENDIF
			\ENDIF	\STATE \}}
	\end{algorithmic}
\end{algorithm}
 \fi

\section{Analysis of Auction Mechanism}\label{sec2}

\subsection{Truthfulness of The Batch Algorithm}
%\noindent (1)Solution feasibility:

\begin{Theorem}
	The batch auction in Algorithm \ref{alg:1} that computes resource allocation and payment is truthful. 
\end{Theorem}

\noindent{\em Proof}: In Algorithm \ref{alg:1}, upon the arrival of user $i$ and our posted price mechanism, the payment $P_{i}$ that user $i$ needs to pay to the cloud provider (if its bid is accepted) depends only on the amount of resources that has been allocated and user $i$'s demand. Which means, user {\em i}'s bidding price does not affect its payment. Therefore, leveraging Lemma \ref{myerson}, our online batch auction is truthful. 
\qed 

\subsection{Solution feasibility of The Batch Algorithm}
\begin{Theorem}
	Algorithm \ref{alg:1} computes a feasible solution to ILP (\ref{max1}).
\end{Theorem}\label{theo_feasible}

\noindent{\em Proof}: $x_{i}$ is initialized to $0$ and updated to $1$ only (line 10 in Algorithm $A_{batch}$), so the solution of our algorithm is binary valued, and satisfies constraint (\ref{m1e}). Container scheduling algorithms $A_{sub1}$ and $A_{sub2}$ guarantee that the schedule {\em S} for each user's bid satisfies constraints (\ref{m1a}), (\ref{m1b}), (\ref{m1f}) and (\ref{m1c}). For container provisioning and scheduling, both $A_{sub1}$ and $A_{sub2}$ select time slots satisfying resource capacity limits, $f^{S}_{ir}(t)+w_{r}(t) \leq \mathbb{C}_{r},  \exists t\in[T].$ Hence constraints (\ref{m1d}) is satisfied. In summary, the solution we obtain is feasible for ILP (\ref{max1}). 
\qed

\begin{Theorem}
	The computational complexity of Batch Algorithm \ref{alg:1} to ILP(\ref{max1}) is polynomial time.
\end{Theorem}\label{theo_complexity}

\noindent{\em Proof}: 

We first consider the case of service chains ($A_{sub} = A_{sub1}$).
Line 1 in Algorithm \ref{alg:1} takes linear time to initialize the price function, primal and dual variables. According to user arrivals, the {\tt while} loop iterates $\rho_{q}$ times to find user $\mu$ with maximum unit resource value, then updates the primal and dual variables in linear time. In the {\tt for} loop (lines 4-6), Algorithm $A_{sub1}$ iterates $\rho_{q}^{2}$ times to select the best schedule of users with maximum utility. Then each $A_{sub1}$ in Algorithm \ref{alg:4} takes $\eta = (d_{i}-t_{s}-\sum_{m\in[M]}N_{im})^{2}$ steps to compute the price of each time slot and examine resource capacity limits for each container. Thus it takes {\em O($M \eta^{2}$}) to choose the utility maximization schedule for user {\em i}. In summary, the running time of $ A_{batch}$ with $A_{sub1}$ is $O(M\eta^{2}\rho_{q}^{2})$. %({\bf double check}). 
We next consider the case of general container graphs ($A_{sub} = A_{sub2}$). The complexity of $A_{sub2}$ is exponential to the number of containers in the container graph, which is mostly small and an be viewed as a constant.  \qed

\subsection{Competitive Ratio of the Batch Algorithm}
%Next we analyse the competitive ratio of our batch algorithm, through a primal-dual analysis framework. 
The {\em competitive ratio} is an upper-bound ratio of the optimal social welfare achieved by ILP (\ref{max1}) to the social welfare achieved by our batch algorithm. The primal-dual framework in our batch algorithm design enables a competitive ratio analysis based on LP duality theory \cite{primaldual}. 
Let $P_{i}$ and $D_{i}$ be the primal objective value (\ref{max2}) and dual objective value (\ref{min}) after accepting user {\em i}'s job, respectively. Then we let $P_{0}$ and $D_{0}$ be the initial objective values of primal (\ref{max2}) and dual (\ref{min}) programs, and $P_{0}$ = 0. $P_{I}$ and $D_{I}$ are the final primal and dual objective values achieved by our algorithm $A_{batch}$. Let {\small $OPT_{1}$} and {\small $OPT_{2}$} be the optimal objective values of (\ref{max1}) and (\ref{max2}), respectively. Since the compact exponential ILP is equivalent to the original ILP, we have {\small $OPT_{1}$ = $OPT_{2}$}, which is hereafter referred to as {\small $OPT$}. %Without loss of generality, if we have $P_{I} \geq \frac{1}{\alpha}D_{I}-\beta$, our algorithm is $\alpha$-competitive, due to weak LP duality, V(algorithm) = $P_{I} \geq \frac{1}{\alpha}D_{I}-\beta \geq \frac{1}{\alpha}OPT-\beta$. 

%The waiting time slots $\theta$ of our batch auction mechanism is within [$\overline{\theta} - Z_{\frac{1}{2}}\sqrt{(\sigma_{1}^{2}+\sigma_{2}^{2})/ \overline{\rho}T}$, $\overline{\theta} + Z_{\frac{1}{2}}\sqrt{(\sigma_{1}^{2}+\sigma_{2}^{2})/ \overline{\rho}T}$]

%For the rationality of our algorithm and discussion of occupation rate of resources $\sigma_{r}$, the job loss rate is less than ten per one hundred jobs for once batch processing ({\bf revise}). Thus we use confidence interval to bound waiting time $\theta$ ({\bf what is waiting time? defined?}).

%We have $\theta \sim N(\mu_{1}-\mu_{2}, \sigma_{1}^{2}+\sigma_{2}^{2})$, $\overline{\theta} \sim N(\mu_{1}-\mu_{2},\frac{\sigma_{1}^{2}+\sigma_{2}^{2}}{\overline{\rho}T} )$. Then
%\begin{center}
%$Z = \frac{\overline{\theta}-(\mu_{1}-\mu_{2})}{\sqrt{(\sigma_{1}^{2}+\sigma_{2}^{2})/\overline{\rho}T}}\sim N(0,1)$

%$P\{\mid \frac{\overline{\theta}-(\mu_{1}-\mu_{2})}{\sqrt{(\sigma_{1}^{2}+\sigma_{2}^{2})/ \overline{\rho}T}} \mid \leq Z_{\frac{1}{2}}\} $= 1-10\%.
%\end{center}

%we can obtain the confidence interval of $\overline{\theta}$ with confidence level over 90 percent, [$\overline{\theta} - Z_{\frac{1}{2}}\sqrt{(\sigma_{1}^{2}+\sigma_{2}^{2})/ \overline{\rho}T}$, $\overline{\theta} + Z_{\frac{1}{2}}\sqrt{(\sigma_{1}^{2}+\sigma_{2}^{2})/ \overline{\rho}T}$].

%If and only if our waiting time $\theta$ lies within the range of confidence interval, the occupation rate of resources $\sigma$ will be satisfied.

\begin{Lemma}\label{D0}
	According to the initial marginal price of each time slot, the initial dual objective value $D_{0}$ is at most $\frac{1}{k}${\em OPT}. 
\end{Lemma}
\noindent {\em Proof:} We first show a lower bound on the optimal social welfare:
{\small
	\begin{center}
		$OPT \geq \sigma \sum_{r\in[R]}\sum_{t\in[T]}F_{r}\mathbb{C}_{r}$.
\end{center}}
Recall that we let $\sigma$ denote the minimum resource occupation rate within slots {\em T}. $F_{r}$ can be interpreted as the minimum social welfare generated by a job per unit of type-{\em r} resource and per unit of time. Therefore, $\sigma \sum_{r\in[R]}\sum_{t\in[T]} F_{r}\mathbb{C}_{r}$ is the minimum social welfare generated by all users.

According to dual (\ref{min}) and marginal price function (\ref{equ:k}):
{\small
	\begin{equation*}
	D_{0} = \sum\nolimits_{t\in [T]} \sum\nolimits_{r\in [R]}\mathbb{C}_{r}\kappa_{r}(0)
	= \sum\nolimits_{t\in [T]} \sum\nolimits_{r\in [R]}\mathbb{C}_{r}(\frac{\sigma F_{r}}{k})
	\end{equation*}
	\begin{equation*}
	= \frac{1}{k}\sum\nolimits_{t\in [T]}\sum\nolimits_{r\in [R]}F_{r}\mathbb{C}_{r} \sigma \leq  \frac{1}{k}OPT 
	\end{equation*}}
Therefore, the the initial dual objective value $D_{0}$ is bounded by  $\frac{1}{k}${\em OPT}.
\qed

\,
\begin{Lemma}\label{lem:1}
	If there is a constant $\alpha > 1$, and the primal and dual objective values increased by handling each user {\em i}'s job satisfy $P_{i}-P_{i-1} \geq \frac{1}{\alpha}(D_{i}-D_{i-1})$,  then the batch algorithm is $\frac{k}{k-1}\alpha$-competitive. 
\end{Lemma}

\,
\noindent {\em Proof:} Since the inequality is satisfied for all users, we sum up the inequality of each user {\em i}:  
\begin{center}
	$P_{I} = \sum_{i}(P_{i}-P_{i-1}) \geq \frac{1}{\alpha}\sum_{i}(D_{i}-D_{i-1}) = \frac{1}{\alpha}(D_{I}-D_{0}).$
\end{center}
According to weak duality and Lemma \ref{D0},  $D_{I} \geq OPT$  and  $D_{0}\geq \frac{1}{k}${\em OPT}. Therefore,  
\begin{center}
	$P_{I}\geq\frac{k-1}{k\alpha}OPT_{1}= \frac{k-1}{k\alpha}OPT_{2},$
\end{center} with the fact that  $P_{0} = 0$. Our batch algorithm is $\frac{k}{k-1}\alpha$-competitive.
\qed 

\vspace*{1.6mm}
Next we will define an {\em Allocation Price Relation} to identify this $\alpha$. If the Allocation Price Relation is satisfied by $\alpha$, the objective values achieved by our algorithm $A_{batch}$ guarantee the inequality in Lemma \ref{lem:1}.

\begin{Definition}
	The Allocation Price Relation for $\alpha \geq 1$ is that $\kappa_{r}^{i-1}(t)(w_{r}^{i}(t)-w_{r}^{i-1}(t))\geq \frac{1}{\alpha}\mathbb{C}_{r}(\kappa_{r}^{i}(t)-\kappa_{r}^{i-1}(t)), \forall i \in [I],\forall r\in[R],\forall t \in [s]$, where $\kappa_{r}^{i}(t)$ represents the price of type-$r$ resource after processing user {\em i}'s job. $w_{r}^{i}(t)$ is the total amount of allocated type-$r$ resource after accepting user {\em i}.
\end{Definition}
\,
\begin{Lemma}For a given $\alpha \geq 1$, if the price function $\kappa_{r}(t)$ satisfies  $\kappa_{r}^{i-1}(t)(w_{r}^{i}(t)-w_{r}^{i-1}(t))\geq \frac{1}{\alpha}\mathbb{C}_{r}(\kappa_{r}^{i}(t)-\kappa_{r}^{i-1}(t)), \forall i \in [I],\forall r\in[R],\forall t \in [l]$, then Algorithm $A_{batch}$  $P_{i}-P_{i-1} \geq \frac{1}{\alpha}(D_{i}-D_{i-1}), \forall i \in [I]$.
\end{Lemma}
\,
\noindent{\em Proof:} If bid {\em i} is rejected,  $P_{i}-P_{i-1} = D_{i}-D_{i-1} = 0$. Then we assume that bid {\em i} is accepted and let $s$ be the job schedule of user $i$. Knowing that our algorithm accepts a bid when constraint (\ref{m3a}) is tight,  $B_{is} = u_{i} + \sum_{r \in[R]} \sum_{ t \in s} f^{s}_{ir}(t) \kappa_{r}^{i-1}(t)$. So the increase of primal objective is:

%\vspace*{-3mm}

\begin{center}
	$P_{i} -P_{i-1} = u_{i} + \sum_{r \in[R]} \sum_{ t \in s} \kappa_{r}^{i-1}(t)(w_{r}^{i}(t)-w_{r}^{i-1}(t))$
\end{center}

%The above inequality holds since our algorithm accepts a job when constraint (\ref{m3a}) is tight, and
According to dual (\ref{min}), the increase of dual objective is:

\begin{center}
	$D_{i} -D_{i-1} = u_{i} + \sum_{r \in[R]}\sum_{ t \in s}\mathbb{C}_{r}(\kappa_{r}^{i}(t)-\kappa_{r}^{i-1}(t))$
\end{center}

Since we have $u_{i}\geq0$, $\alpha \geq 1$ and $ \kappa_{r}^{i-1}(t)(w_{r}^{i}(t)-w_{r}^{i-1}(t))\geq \frac{1}{\alpha}\mathbb{C}_{r}(\kappa_{r}^{i}(t)-\kappa_{r}^{i-1}(t))$:
{\small \begin{equation*}
	\begin{aligned}
	P_{i} -P_{i-1}& = u_{i} + \sum_{r \in[R]} \sum_{ t \in s} \kappa_{r}^{i-1}(t)(w_{r}^{i}(t)-w_{r}^{i-1}(t))\\
	&\geq u_{i} + \frac{1}{\alpha} \sum\nolimits_{r \in[R]}\sum\nolimits_{ t \in s}\mathbb{C}_{r}(\kappa_{r}^{i}(t)-\kappa_{r}^{i-1}(t))\\
	& \geq \frac{1}{\alpha}(u_{i} + \sum\nolimits_{r \in[R]}\sum\nolimits_{ t \in s}\mathbb{C}_{r}(\kappa_{r}^{i}(t)-\kappa_{r}^{i-1}(t)))\\
	&=\frac{1}{\alpha}(D_{i}-D_{i-1})\\ 
	\end{aligned}
	\end{equation*}\qed}

%Since we have $u_{i}\geq0$ and $\alpha \geq 1$, $P_{i}-P_{i-1} \geq \frac{1}{\alpha}(D_{i}-D_{i-1})$ is equivalent to $\kappa_{r}^{i-1}(t)(w_{r}^{i}(t)-w_{r}^{i-1}(t))\geq \frac{1}{\alpha}\mathbb{C}_{r}(\kappa_{r}^{i}(t)-\kappa_{r}^{i-1}(t))$. 

We next try to find the $\alpha_{r}$ for type-{\em r} resource that satisfies the Allocation Price Relationship. Thus the $\alpha$ is the maximum value among all $\alpha_{r}$. Since the capacity of type-{\em r} resource is larger than a user demand, we let $dw_{r}(t)$ denote $w_{r}^{i}(t)-w_{r}^{i-1}(t)$. We first prepare with the following definition.

\begin{Definition}
	The Differential Allocation Price Relation for $A_{batch}$ with a given parameter  $\alpha_{r}\geq 1$  $\kappa_{r}(t)dw_{r}(t)\geq \frac{1}{\alpha_{r}}\mathbb{C}_{r}d\kappa_{r}(t), \forall i\in[I], \forall r\in[R], \forall t\in[s]$.
\end{Definition}
\,
\begin{Lemma}\label{lemma_alpha}
	The marginal price defined in (5) satisfies the Differential Allocation Price Relation, and we can get $\alpha_{r}= ln(\frac{k D_{r}}{\sigma F_{r}})$.
\end{Lemma}

\noindent {\em Proof:} The derivative of the marginal price function is:
{\small
	\begin{equation*}
	d\kappa_{r}(t) = \kappa^{'}_{r}(w_{r}(t))dw_{r}(t)
	= \frac{\sigma F_{r}}{k}(\frac{k D_{r}}{\sigma F_{r}})^\frac{w_{r}(t)}{\mathbb{C}_{r}}\frac{1}{\mathbb{C}_{r}}ln(\frac{kD_{r}}{\sigma  F_{r}})dw_{r}(t).
	\end{equation*}}
{\small 
	\begin{equation*}
	\textrm{Therefore:}\quad \frac{\sigma F_{r}}{ k}(\frac{kD_{r}}{\sigma F_{r}})^\frac{w_{r}(t)}{\mathbb{C}_{r}} \geq  \frac{\mathbb{C}_{r}}{\alpha_{r}}\frac{\sigma F_{r}}{k}(\frac{k D_{r}}{\sigma F_{r}})^\frac{w_{r}(t)}{\mathbb{C}_{r}}\frac{1}{\mathbb{C}_{r}}ln(\frac{k D_{r}}{\sigma F_{r}})
	\end{equation*}
	\begin{equation*}
	\geq  \frac{1}{\alpha_{r}}(\frac{\sigma F_{r}}{k}(\frac{k D_{r}}{\sigma F_{r}})^\frac{w_{r}(t)}{\mathbb{C}_{r}}ln(\frac{k D_{r}}{\sigma F_{r}})),
	\Rightarrow  \alpha_{r} \geq  ln(\frac{k D_{r}}{\sigma F_{r}})
	\end{equation*} }

\noindent Thus we can obtain $\alpha_{r}= ln(\frac{k D_{r}}{\sigma F_{r}})$.
\qed

\begin{Lemma}\label{the_comp}
	The batch auction Algorithm  $A_{batch}$ is $\frac{k}{k-1}\alpha$-competitive in social welfare with $\alpha = max_{r\in[R]}ln(\frac{k D_{r}}{\sigma F_{r}})$. %({\bf do we have multiple definitions of $\alpha$ in this paper?}).
\end{Lemma}

\noindent{\em Proof:}  Lemma \ref{lemma_alpha} implies that $\alpha = max_{r\in[R]}ln(\frac{k D_{r}}{\sigma F_{r}}) $ satisfies the Differential Allocation Price Relation of all kinds of resources. Since the above mentioned, 
$dw_{r}(t) = w_{r}^{i}(t)-w_{r}^{i-1}(t)$,
{\small
	\begin{equation*}
	\begin{aligned}
	d\kappa_{r}(t) &= \kappa_{r}^{'}(w_{r}(t))dw_{r}(t)  =\kappa_{r}^{'}(w_{r}(t))(w_{r}^{i}(t)-w_{r}^{i-1}(t)) \\
	&= \kappa_{r}^{i}(t) - \kappa_{r}^{i-1}(t). 
	\end{aligned}
	\end{equation*}}
Thus, we can obtain  $\alpha = max_{r\in[R]}ln(\frac{k D_{r}}{\sigma F_{r}})$ due to the Allocation Price Relationship. \qed

\begin{Theorem}\label{lem4}
	If {\em k} satisfies $k-1 = \max_{r\in[R]}ln(\frac{k D_{r}}{\sigma F_{r}})$ and $k>1 $, the competitive ratio of batch auction algorithm is minimum, and is equal to {\em k}.
\end{Theorem}

\noindent{\em Proof:} We assumpt that $\varpi = max_{r\in[R]}(\frac{D_{r}}{\sigma F_{r}})$. By Lemma \ref{lem4}, the competitive ratio of our batch algorithm is $\frac{k}{k-1}\alpha =  \frac{k}{k-1}ln(\frac{k D_{r^{*}}}{\sigma   F_{r^{*}}})$ =$\frac{k}{k-1}ln(k\varpi) $, thus the competitive ratio is a function of {\em k}. Differentiating  $\frac{k}{k-1}ln(k\varpi) $ on {\em k} is:
{\small
	$$(\frac{k}{k-1}ln(k\varpi) )^{'} =\frac{k-1-ln(k\varpi)}{(k-1)^{2}} $$}
It suffices to show that $(\frac{k-1-ln(k\varpi)}{(k-1)^{2}} )^{'}$ is positive as $k \in [1,\varpropto]$. When k satisfies $k-1 =ln(k\varpi) $ and $k>1$, we can obtain the minimum competitive ratio:{\small
	$$ \frac{k}{k-1}ln(k\varpi) =\frac{k}{ln(k\varpi)}ln(k\varpi) =k.$$}

\begin{figure}[!htbp]
	\begin {center}
	\hspace*{-7mm}
	\includegraphics[width=0.45\textwidth]{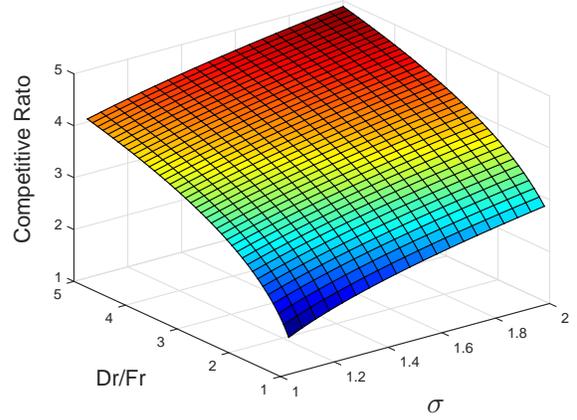}	
	\vspace{-1mm}
	\caption{Theoretical Competitive Ratio.}
	\label{fig:3d}
	\end {center}
	\vspace*{-3mm}
\end{figure}

If we consider the case that competition for resource is intense, the $\sigma$ is close to 1. When $D_{r}/F_{r}$ is {\em 2}, the competitive ratio is close to $2.85$, as illustrated in Fig. \ref{fig:3d}. \qed 

\subsection{Setting The Batch Interval $\theta$}\label{theta}

In our batch auction, the more jobs we handle in a batch, the more information we have for social welfare maximization. Nonetheless, we can't over-extend the length of a batch given that cloud jobs have deadlines to meet. Precise optimization of the job interval length is left as future research, and we provide here a brief discussion only. Let $W_{i}$ be the time required to execute a job $i$, and $\overline{\rho}$ be the expected number of user arrivals per slot. In general, an appropriate length of a batch round depends on values of $W_{i}$, deadline $d_{i}$ and arrival time $t_{i}$ of user {\em i}, $i\in [I]$. We can set a target threshold on the job loss rate ({\em e.g.}, 10\%), the ratio of jobs who cannot meet their deadlines due to delayed bid admission.

%According to the traces in Google cluster data \cite{Googleclusterdata}, 
Assume that job processing time and $d_{i}-t_{i}$ are normally distributed, by $N(a_{1}, b_{1}^{2})$ and $N(a_{2}, b_{2}^{2})$, respectively. The max waiting time for each user equals $d_{i}-t_{i}-W_{i}$, and is also normally distributed as $N(a_{1}-a_{2}, b_{1}^{2}+b_{2}^{2})$. If user {\em i}'s maximum waiting time $\theta_{i} < \theta$, we will lose this job. Thus the length of batch interval $\theta$ can be set by (for $\le$ 10\% job loss):

\begin{center}
	\{$\max \theta,$ $ s.t. \sum_{t=1}^{\theta}F(\theta)\leq 0.1, \theta \in \{1, 2, 3, 4,...\} .  $\}
\end{center}

Where $F(\theta)$ is the Normal cumulative  distribution function of $\theta$.

\begin{figure*}[!htbp] 
	\hspace{1mm}
	\begin{minipage}{0.33\textwidth}
		\begin {center}
		\includegraphics[width=1\textwidth]{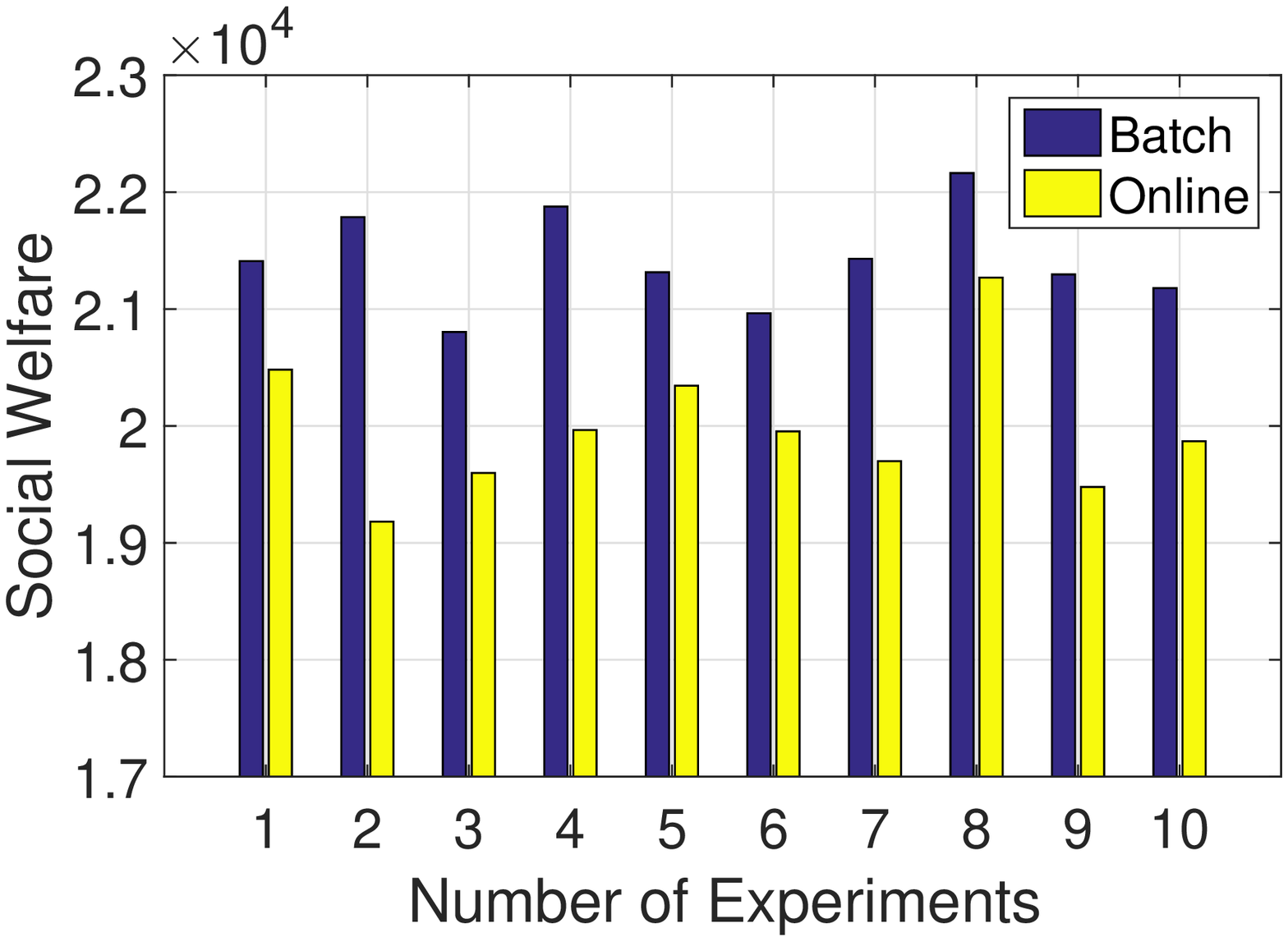}	
		\vspace*{-5mm}		
		\setlength{\abovecaptionskip}{10pt}\caption{Social welfare, batch vs. online auctions.}
		\label{fig:batch_online}
		\end {center}	
		\vspace*{-2mm}
	\end{minipage}
	%	\hspace{1mm}
	\begin{minipage}{0.33\textwidth}
		\begin {center}
		\includegraphics[width=1\textwidth]{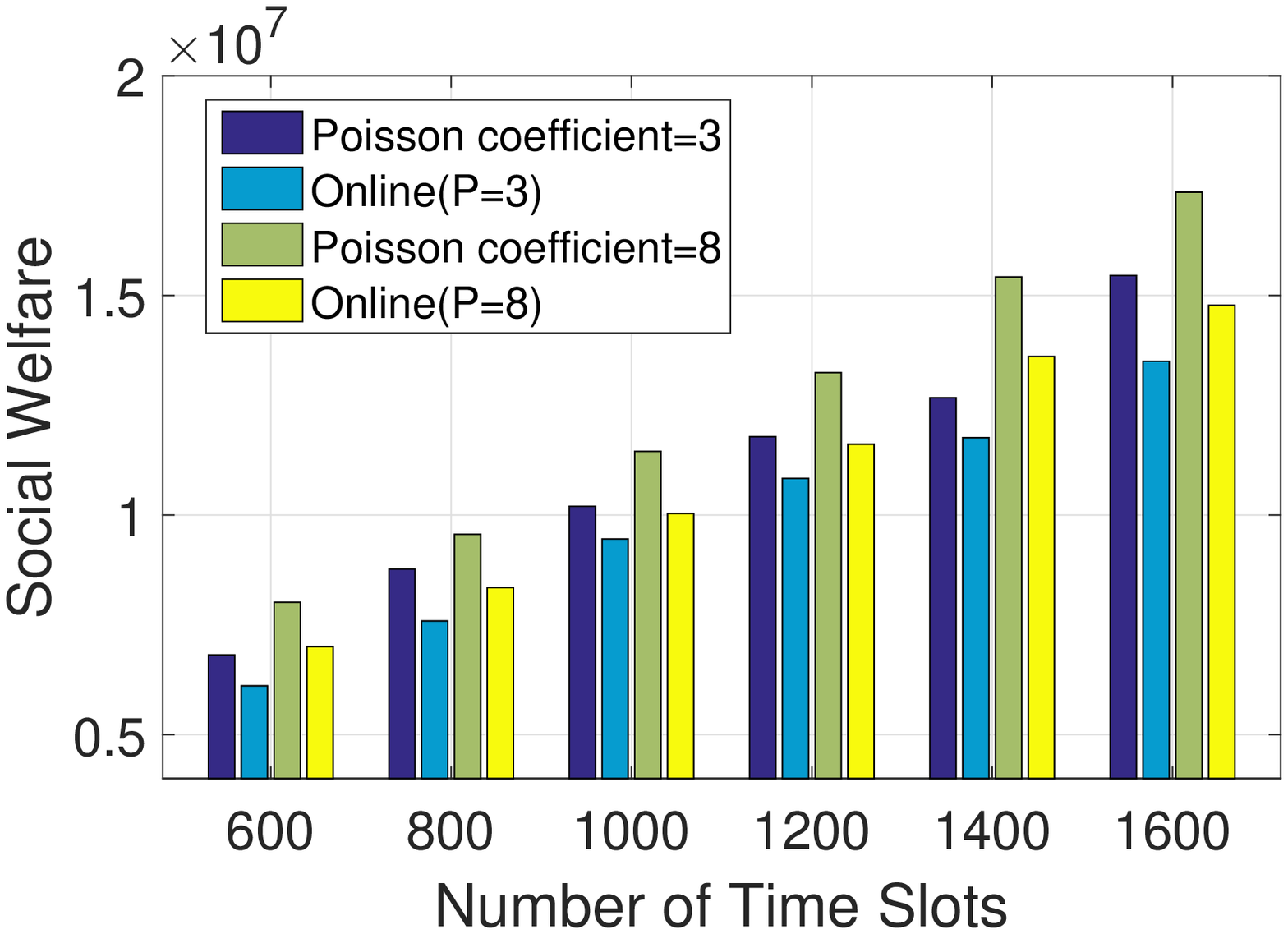}	
		\vspace*{-5mm}
		\setlength{\abovecaptionskip}{10pt}\caption{Social Welfare achieved by $A_{batch}$.}
		\label{fig:interval}
		\end {center}	
		\vspace*{2mm}
	\end{minipage}
	\hspace{1mm} 
	\begin{minipage}{0.33\textwidth}
		\begin {center}
		\includegraphics[width=1\textwidth]{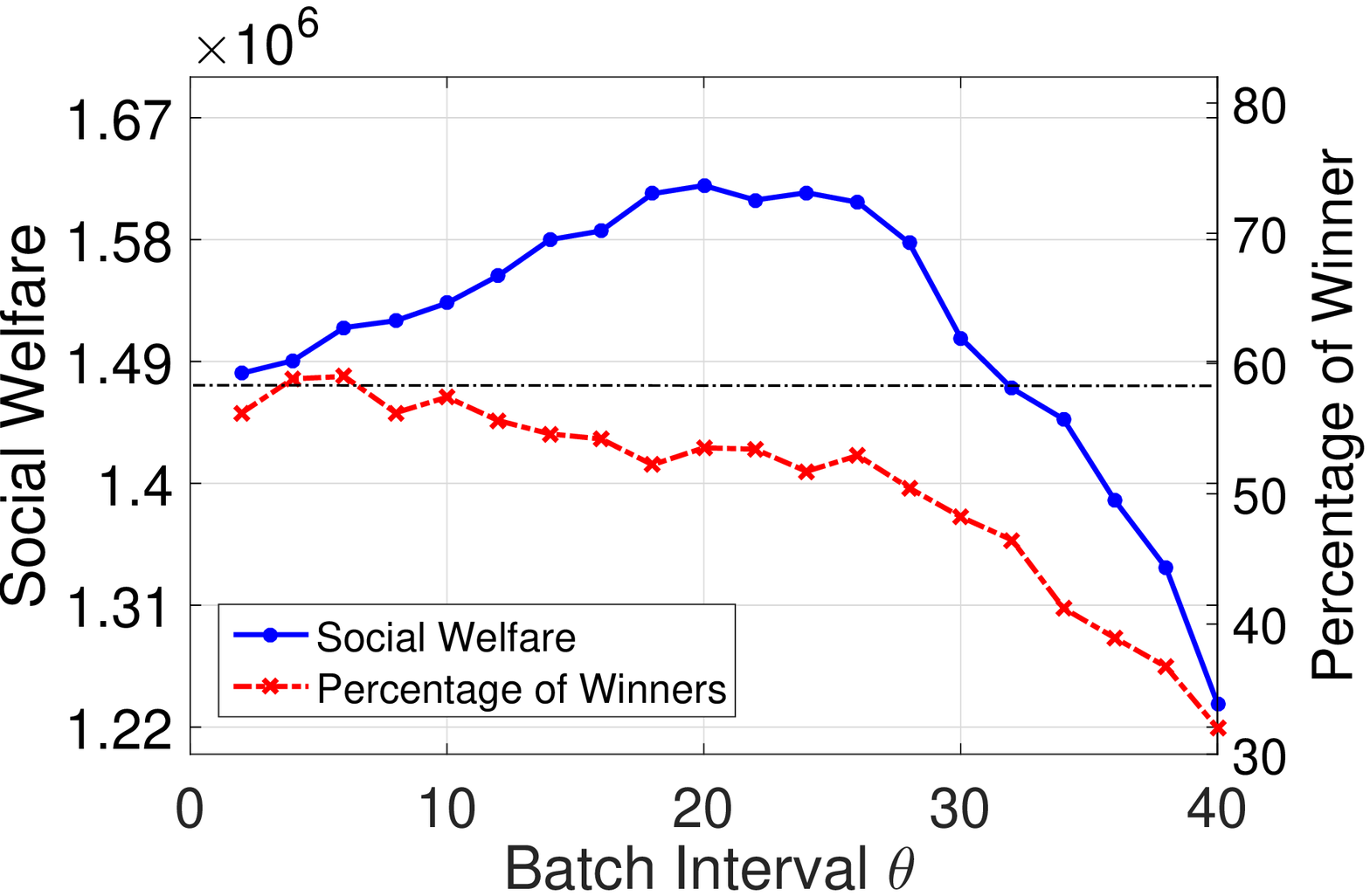}			\vspace*{-6mm}		
		\setlength{\abovecaptionskip}{10pt}\caption{Social Welfare and percentage of winners, varying batch length.}
		\label{fig:welfare_interval}
		\end {center}	
		\vspace*{-4mm}
	\end{minipage}%
	%	
	%\hspace{1mm} 
	\vspace*{-4mm}
\end{figure*}

\begin{figure*}[!htbp] 
	\hspace{1mm}
	\begin{minipage}{0.33\textwidth}
		\begin {center}
		\includegraphics[width=1\textwidth]{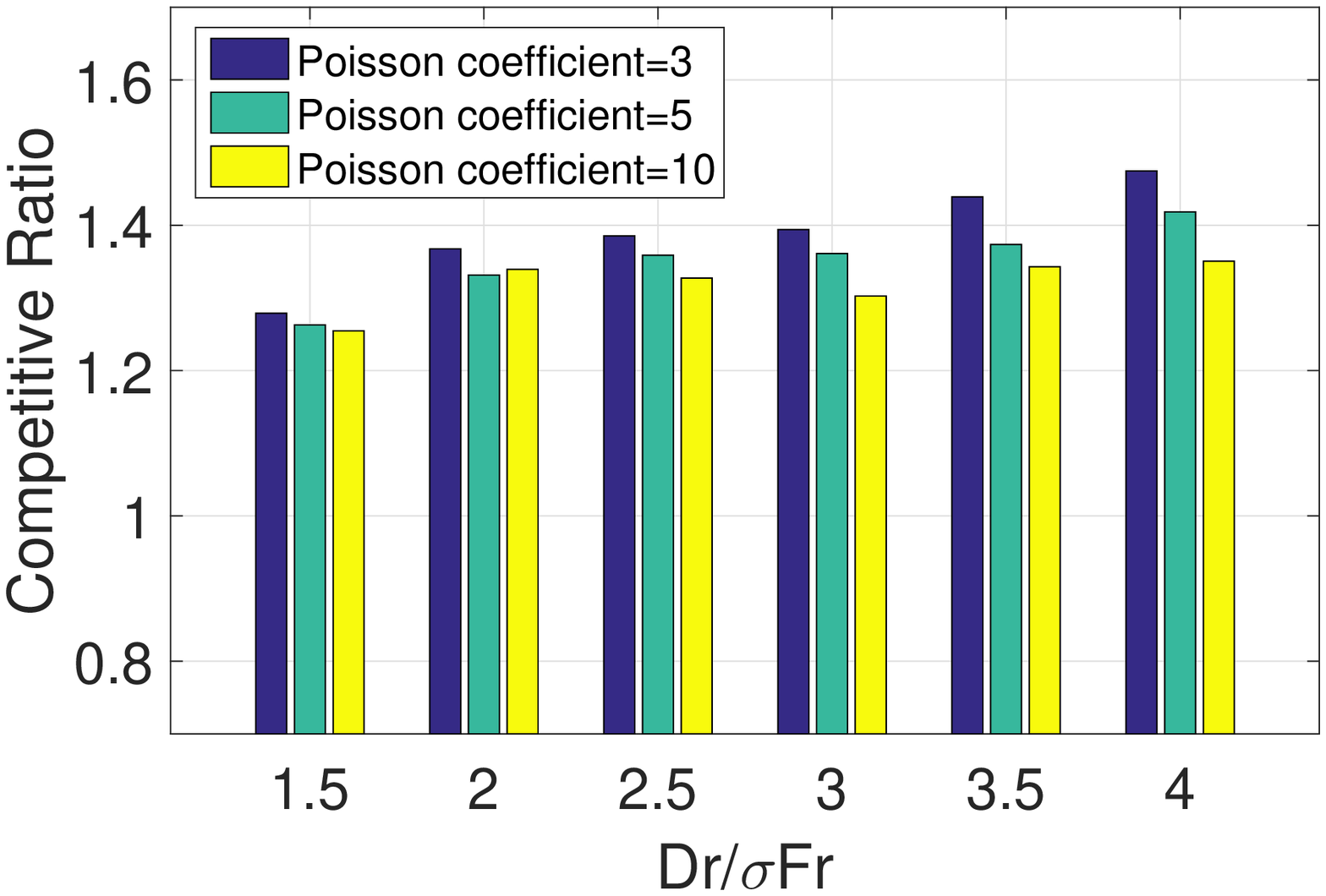}	
		\vspace*{-5mm}		
		\setlength{\abovecaptionskip}{10pt}\caption{Competitive ratio of auction algorithm $A_{batch}$.}
		\label{fig:ratio_UL}
		\end {center}	
		\vspace*{-2mm}
	\end{minipage}
	%	\hspace{1mm}
	\begin{minipage}{0.33\textwidth}
		\begin {center}
		\includegraphics[width=1\textwidth]{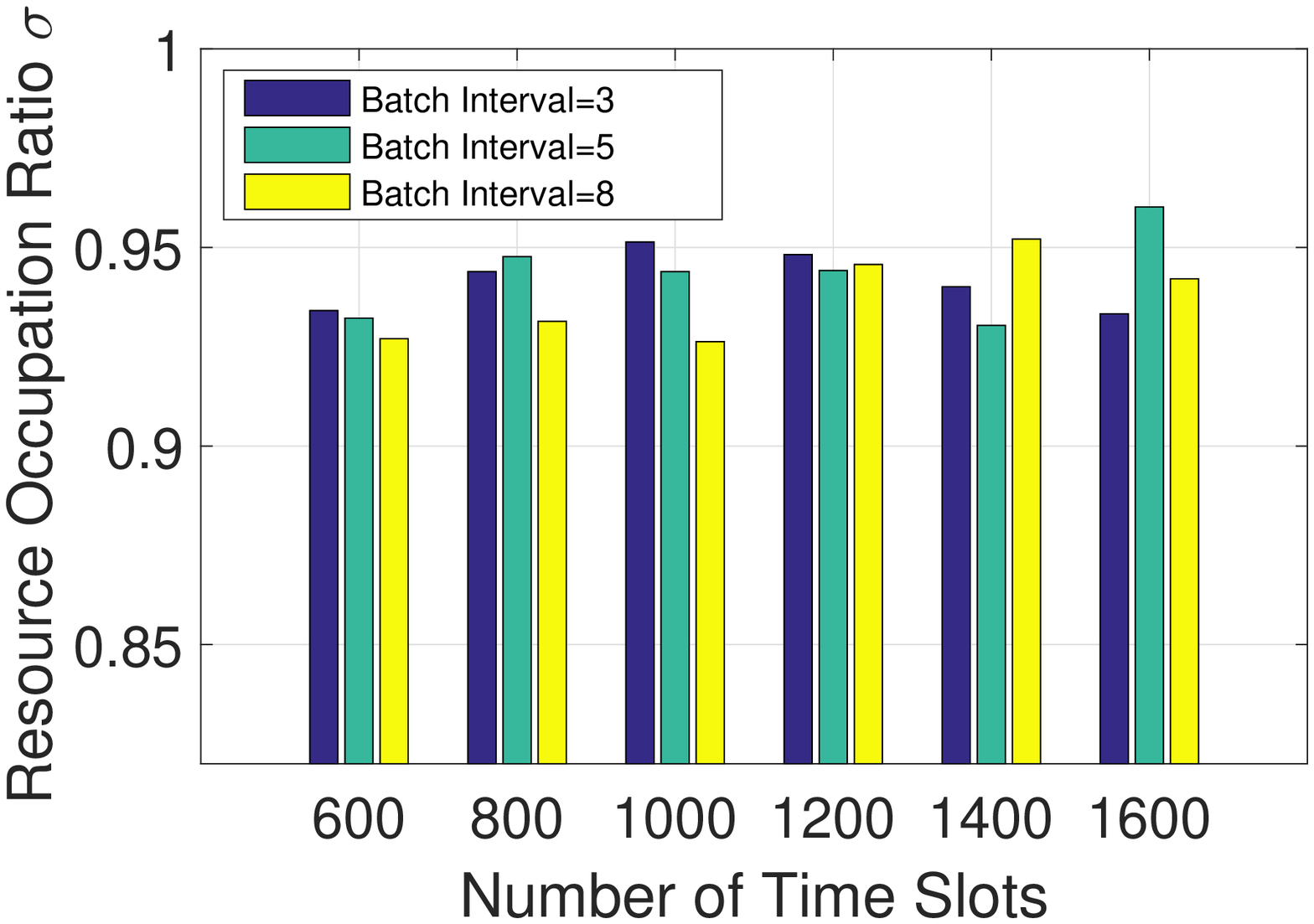}	
		\vspace*{-5mm}
		\setlength{\abovecaptionskip}{10pt}\caption{Resource occupation ratio $\sigma$.}
		\label{occupation}
		\end {center}	
		\vspace*{2mm}
	\end{minipage}
	\hspace{1mm} 
	\begin{minipage}{0.33\textwidth}
		\begin {center}
		\includegraphics[width=1\textwidth]{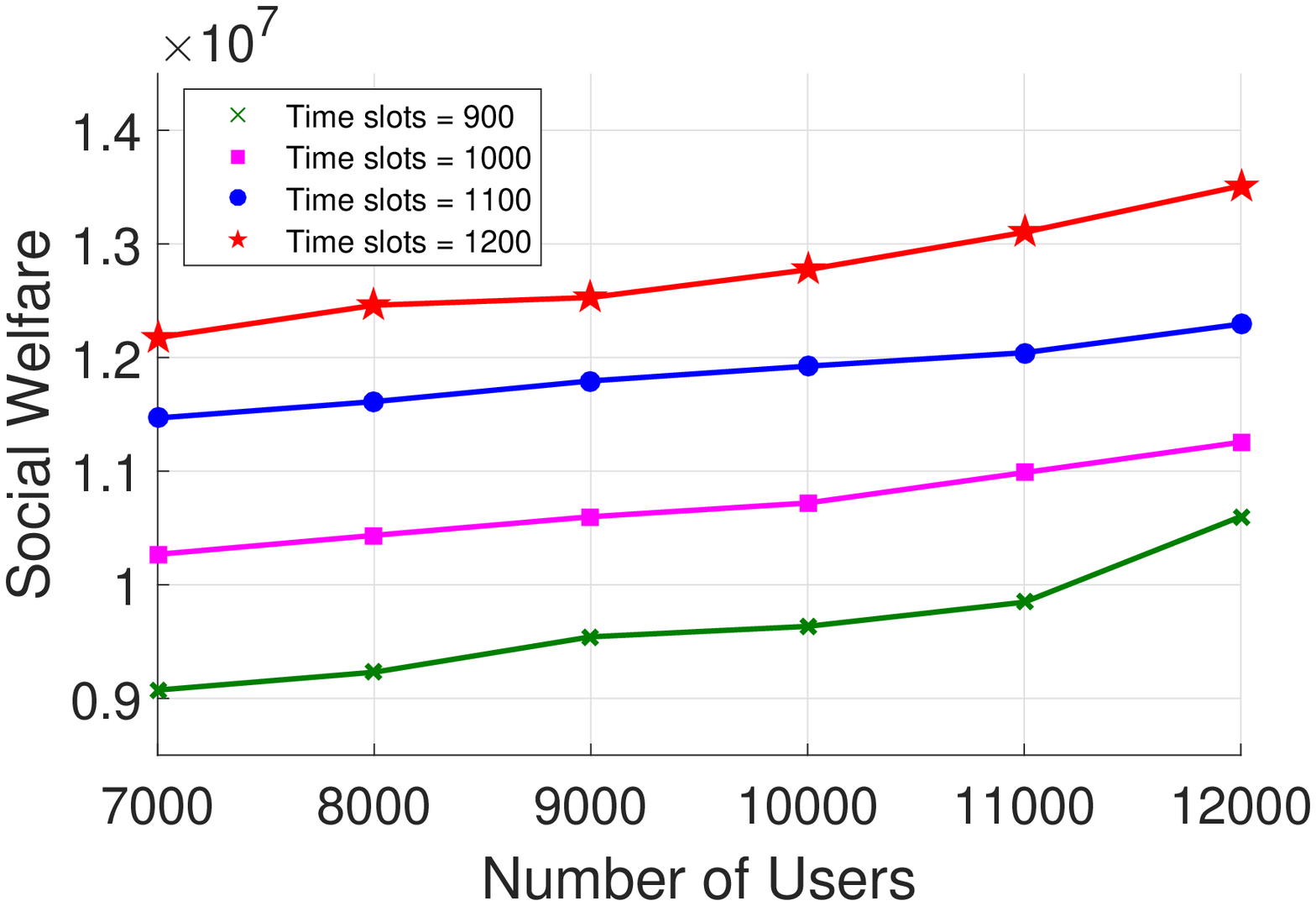}			\vspace*{-6mm}		
		\setlength{\abovecaptionskip}{10pt}\caption{Social Welfare of $A_{batch}$, varying user population.}
		\label{wel-user}
		\end {center}	
		\vspace*{-4mm}
	\end{minipage}%
	%	
	%\hspace{1mm} 
	\vspace*{-4mm}
\end{figure*}

\section{PERFORMANCE EVALUATION}
\label{evaluation}
We evaluate our batch auction algorithm $A_{batch}$ and its sub-algorithms by trace-driven simulation studies. We leverage Google cluster data \cite{Googleclusterdata}, which captures rich information on user jobs, including start time, resource demand (CPU, RAM and Disk), and duration. We translate cloud job requests into bids, arriving in a one month time window. We assume that each sub-task consumes [1,10] slots, and each time slot is one hour. Job deadlines are set randomly between the arrival time and system end time. %Our batch auction waiting various number of time slots once per batch, during these times, comparing the objective of auction under the different number of arrival users, and considering the influence of $D_{r}/\sigma F_{r}$ to social welfare. 
The demand of resources (CPU, RAM and Disk) is set randomly between [0, 1], with the resource capacity set to 50. We use {\em user density} to express the number of users in one batch interval, arriving as a Poisson process. 

%Next we prove the performance of our batch algorithm $A_{batch}$ by means of competitive ratio, social welfare and users satisfaction. 

\vspace{1mm}
\noindent {\em A.~Comparison with Classic Online Auctions}

We compare our batch auction with a traditional online auction in terms of social welfare, as shown in Fig.~\ref{fig:batch_online}. Under the same simulation settings, we compare the two algorithms in 10 different sets of simulation studies. Our batch auction achieves a higher social welfare in all of them. Intuitively, the online auction processes bids in a FCFS fashion, while the batch auction considers most attractive bids first in each batch.  Fig.~\ref{fig:interval} shows another set of comparisons. The superiority of batch auction remains clear, with different number of time slots and user density. Social welfare fluctuates with the increase of the number of users and user density. The batch auction performs better with higher user density. 
The influence of different batch interval $\theta$ for the batch performance is illustrated in Fig.~\ref{fig:welfare_interval}. As $\theta$ grows, the cloud social welfare initially grows as well. However, when $\theta$ is too large so that more bids are lost due to delays, as we can see in Fig.~\ref{fig:welfare_interval}, a gradual decrease in the percentage of winners leads to a decreasing trend in social welfare. Recall that in the analysis of $\theta$ in the previous section, a too large $\theta$ is not suitable for our batch auction.

\vspace{1mm}
\noindent {\em B.~Competitive Ratio of The Batch auction}

Next we study the competitive ratio achieved by our batch auction. 
As we proved in Theorem \ref{the_comp}, the competitive ratio depends on $D_{r}/\sigma F_{r}$. Fig.~\ref{fig:ratio_UL} shows that the competitive ratio grows as $D_{r}/\sigma F_{r}$ increases. The observed competitive ratio is much better than the theoretical bound and remains smaller than 2; this can be partly explained by the fact that the theoretical bound is a pessimistic worst case scenario uncommon in practice. The ratio fluctuates with user population and sightly decreases with as $D_{r}/\sigma F_{r}$ decreases. The batch auction favors intensive user arrivals.

\vspace{1mm}
\noindent {\em C.~Performance of $A_{batch}$: The Role of System Parameters}

We next examine the resource occupation ratio $\sigma$ (defined in Sec.~\ref{model}) of our batch auction. As we can see in Fig.~\ref{occupation}, under different numbers of time slots and user density, the resource occupation ratio of the batch auction mechanism is constantly beyond 90\% and often close to 1. Fig.~\ref{wel-user} demonstrates the variation of social welfare with different number of users. The social welfare grows mildly but steadily as the number of users and the number of time slots grow.

\section{CONCLUSION}
\label{sec:conclusion}
This work is the first in the cloud computing literature that studies efficient auction algorithm design for container services. It is also the first that designs batch online auctions, aiming at more informed decision making through exploiting the elastic nature of cloud jobs. We combined techniques from compact exponential optimization, posted price mechanisms, and primal-dual algorithms for designing a cloud container auction that is incentive compatible, computationally efficient, and economically efficient. As future directions, it will be interesting to study (i) cloud jobs that cannot be suspended and resumed; (ii) pre-processing of cloud jobs with tight deadlines to choose between immediate acceptance or delayed processing of their bids; and (iii) cloud container auctions that make revocable decisions, where a partially executed cloud job may or may not contribute towards social welfare of the cloud.

{\small
	\bibliographystyle{IEEEtran}
	\bibliography{reference} \label{reference}
}

\end{document}